\documentclass[12pt]{article}

\usepackage{amsmath}
\usepackage{amssymb}
\usepackage{euscript}

\usepackage{epsfig}

\usepackage{cite}  

\def\CCol{{\tt SANC}}

\begin{document}

\begin{titlepage}
\thispagestyle{empty}

\begin{flushright}
{\bf  
CERN-TH/2002-315\\
} 
\end{flushright}
 
\noindent
\vspace*{5mm}
\begin{center}
{
     \Large\bf Comparison of {\tt SANC} with  {\tt KORALZ} and {\tt PHOTOS}}
\end{center}

\vspace*{2mm} 
\begin{center}
  {\bf A. Andonov$^{a}$,}
  {\bf S. Jadach$^{b}$,}~~
  {\bf G. Nanava$^{a}$ and}~~
  {\bf Z. W\c{a}s$^{b,c}$ }
  \\
  \vspace{5mm}
  {\em $^a$ Lab. of Nuclear Problems, JINR, RU-141980 Dubna, Russia}\\
  \vspace{1mm}
  {\em $^b$ Institute of Nuclear Physics,
           ul. Radzikowskiego 152, PL-31-342 Cracow, Poland}\\
  \vspace{1mm}
  {\em $^c$  CERN, Theory Division, CH-1211 Geneva 23, Switzerland }
\end{center}
\vfill

\vspace*{5mm} 
\begin{abstract}
Using the {\tt SANC} system we study the one-loop electroweak standard model
prediction, including virtual and real photon emissions, for the decays
of on-shell vector and scalar bosons $B \to f {\bar f} (\gamma)$, 
where  $B$ is a vector boson, $Z$ or $W$,
or a Standard Model  Higgs.
The complete one-loop corrections and exact photon emission
matrix element are taken into account. For the 
phase-space integration, the  Monte Carlo technique is used.
For $Z$ decay the QED part of the calculation is first cross-checked  with
the exact one-loop QED prediction of {\tt KORALZ}.
For Higgs boson and $W$ decays, a comparison is made with the
approximate QED calculation of {\tt PHOTOS} Monte Carlo.
This provides a useful element for the evaluation of the theoretical uncertainty
of {\tt PHOTOS},
very  interesting for its application in ongoing LEP2 and future LC and LHC phenomenology. 
\end{abstract}  
\vspace*{2mm}

\begin{center}
  {\em Submitted to Acta Physica Polonica}
\end{center}

\vspace*{1mm}
\vfill

\vspace*{1mm}
\bigskip
\footnoterule
\noindent
{\footnotesize \noindent
Work supported in part by the European Union 5-th Framework under contract HPRN-CT-2000-00149,
Polish State Committee for Scientific Research 
(KBN) grant 2 P03B 001 22, 
NATO grant PST.CLG.977751
and INTAS M$^{o}$ 00-00313.
}
\end{titlepage}

\normalsize
\newpage

\section{Introduction}

The \CCol\ project of Ref.~\cite{Bardin:2002am} has several purposes.
The intermediate  goal is to summarize
and consolidate the effort of the last three decades  in calculating
Standard Model radiative corrections for LEP, in a well organized 
 calculational environment for future reference.
However, it is aimed not only at training young 
researchers and students, but at some remaining
calculational projects for LEP as well.

\CCol\ provides an Internet-oriented, graphic interface 
platform, which is meant to serve as a starting point
for longer term research efforts in the area of the higher order (multi-loop)
calculations within the SM and beyond, for  the experiments
at future high energy colliders.
An important lesson from the LEP experiments \cite{Kobel:2000aw}
is that the desirable
way of providing theoretical predictions is in the form of  Monte Carlo
event generators.
This aspect has been taken into account in the development of \CCol\
from an early stage of its development.

The currently available  version of \CCol\  can construct
one-loop spin amplitudes for the
decays of the gauge bosons $W$ and $Z$ and Higgs boson $H$.
All of the Born and one-loop-corrected spin amplitudes
are generated interactively from  scratch%
\footnote{The {\tt form2} codes related to the review of
 ref.~\cite{Bardin:1999ak} were exploited at the early stage 
 of the \CCol\ development.
}
by \CCol\ with the help of the
algebraic package {\tt Form3} \cite{Vermaseren:2000nd}
in the form of  Fortran77 source codes,
and are then used in  the MC generation/intgration part of the package.
For the moment, \CCol\ features single real photon emission, in
the calculations of the total rate and
decay spectra of the $B \to f {\bar f} (\gamma)$ process.
The complete spin polarization density matrix of the decaying boson
is  taken into account as well.

The integration, with the Monte Carlo method, 
over the three (two)-body final state is done without 
any approximation (in particular small mass approximation is not used).
The program provides MC events with constant weight (unweighted events).
The whole system is, therefore, fairly self-contained and complete%
\footnote{That  is why it may be valuable for teaching and training.}.

It is of the utmost importance for such a new system  to
precisely reproduce known  results.
Concerning virtual corrections,  comparisons were 
performed  with the  calculations of  
refs.~\cite{Bardin:1999yd,Montagna:1998kp}.
The precise comparisons with the {\tt FeynArt} package
\cite{Kublbeck:1990xc}
were made
in~\cite{Andonov:2002rr,Bardin:2000kn,andonov02:updat_cal} as well.

Our paper is organized as follows: in the next section we describe in  detail
the set of observables chosen for tests and
the input parameter set-ups for comparisons \CCol\ with {\tt KORALZ} \cite{Jadach:1994yv}
and  \CCol\ {\tt PHOTOS} \cite{Barberio:1994qi}.
Section 3 is devoted to a discussion of  \CCol\ technical reliability in the domain
of QED bremsstrahlung. To this end,
comparisons with {\tt KORALZ}
and leading-log distributions from  {\tt PHOTOS} are collected.
Section 4 describes  comparisons
of  \CCol{} with {\tt PHOTOS}, focusing on the non-leading
contributions in $W$ and $H$ decays
and on the relevant discussion of {\tt PHOTOS} physical uncertainties.
The summary in section 5 closes the paper.

\section{Initialization set-ups for \CCol\,{},  {\tt KORALZ}
         and {\tt PHOTOS} runs}

In the following sections we compare predictions from the programs
\CCol,  {\tt KORALZ} and {\tt PHOTOS}. It is essential that the initialization  
be identical in all cases and close to the physical reality;
in particular the following options are set in the three programs:
\begin{itemize}
\item
  In \CCol\ we switch off the EW part of the radiative corrections.
  The soft photon limit is kept at 0.005 of the decaying particle mass. 
\item
  In {\tt KORALZ} we switch to the ${\cal O}(\alpha)$ mode of operation
  (no exponentiation) with final state bremsstrahlung only.
  We set the CMS energy equal to the $Z$ mass. 
  The $s$-channel $\gamma$ exchange is switched off and the
  soft photon limit is kept at 0.01 of the ``beam energy''.
\item
  In {\tt PHOTOS} we switch off the double bremsstrahlung corrections
  but we keep interference effects. The soft/hard photon limit is kept at
  0.005 of the decaying particle mass.
  For the generation of the Born-level two-body decays,
   we use the Monte Carlo generation from
  \CCol{}.
\end{itemize}

To visualize the differences (or the agreement) between the calculations,
we choose a certain class of (pseudo-)observables, more precisely 
the one-dimensional distributions, which  are quite similar to the one
used in the  first tests of {\tt PHOTOS} reported in \cite{Barberio:1991ms}. 
To visualize the usually small differences,
we plot ratios of the predictions from two programs rather 
than the distributions themselves.

\vskip 2 mm

List of observables:
\begin{itemize}
\item
{\bf  -A-} {\it Photon energy in the decaying particle rest frame:} this
observable is sensitive mainly to the leading-log (i.e. collinear) non-infrared
(i.e. not soft) component of the distributions.
\item
{\bf  -B-}  {\it Energy of the final state charged particle:} as in the 
previous case this
observable is sensitive mainly to the leading-log (i.e. collinear) non-infrared
(i.e. not soft) component of the distributions.
\item
{\bf  -C-}  {\it Angle of the photon with respect to one of the charged
final state particles:} this
observable is sensitive mainly to the non-collinear  (i.e. non-leading-log) 
but soft 
(i.e. infrared) component of the distributions.
\item
{\bf  -D-}  {\it Acollinearity angle of the final state charged particles:} this
observable is sensitive mainly to the non-collinear  (i.e. non-leading-log) 
and  non-soft 
(i.e. non-infrared) component of the distributions.
\end{itemize}

\section{Technical tests of   \CCol\ }

Let us first cross check  \CCol\ predictions 
in the case of $Z \to \mu^+ \mu^-$ decay and for the real photon emission
with {\tt KORALZ}.
The corresponding
part of the {\tt KORALZ} code, essentially an improved emulation of the 
MC program {\tt MUSTRAL} \cite{Berends:1983mi}, was thorougly tested
over two decades and is an excellent candidate for the benchmark test
for \CCol\ matrix-element and phase-space generation. 
It can be noted that the phase-space parametrization
of the single photon emission in \CCol, although developed independently,
is essentially the same as in 
{\tt KORALB} \cite{Jadach:1991mv,Jadach:1985iy,Jadach:1984ac}.

As we can see from figs. 1 and 2 the 
level of the agreement  for all distributions of the types {\bf A, B, C} and
{\bf D} is satisfactory, at the level 
of better than 0.5\% (or at the level of statistical error).
We have checked that the ${\cal O} (\alpha)$ FSR correction to the 
total decay rate
calculated as a difference of \CCol\ results, with FSR on and off,
agrees well with the standard factor $1 + {3 \over 4} {\alpha \over \pi}= 1.001743$.
In fact we get $1.001733 \pm 2.8\times 10^{-5}$. 
The agreement is thus better than $10^{-4}$. {\it We can thus conclude that \CCol\ is ``commissioned'' for the $Z$ decay.}

\begin{figure}[!ht]  
\setlength{\unitlength}{0.1mm}  
\begin{picture}(1600,800)  
\put( 375,750){\makebox(0,0)[b]{\large }}  
\put(1225,750){\makebox(0,0)[b]{\large }}  
\put( -60, 00){\makebox(0,0)[lb]{\epsfig{file=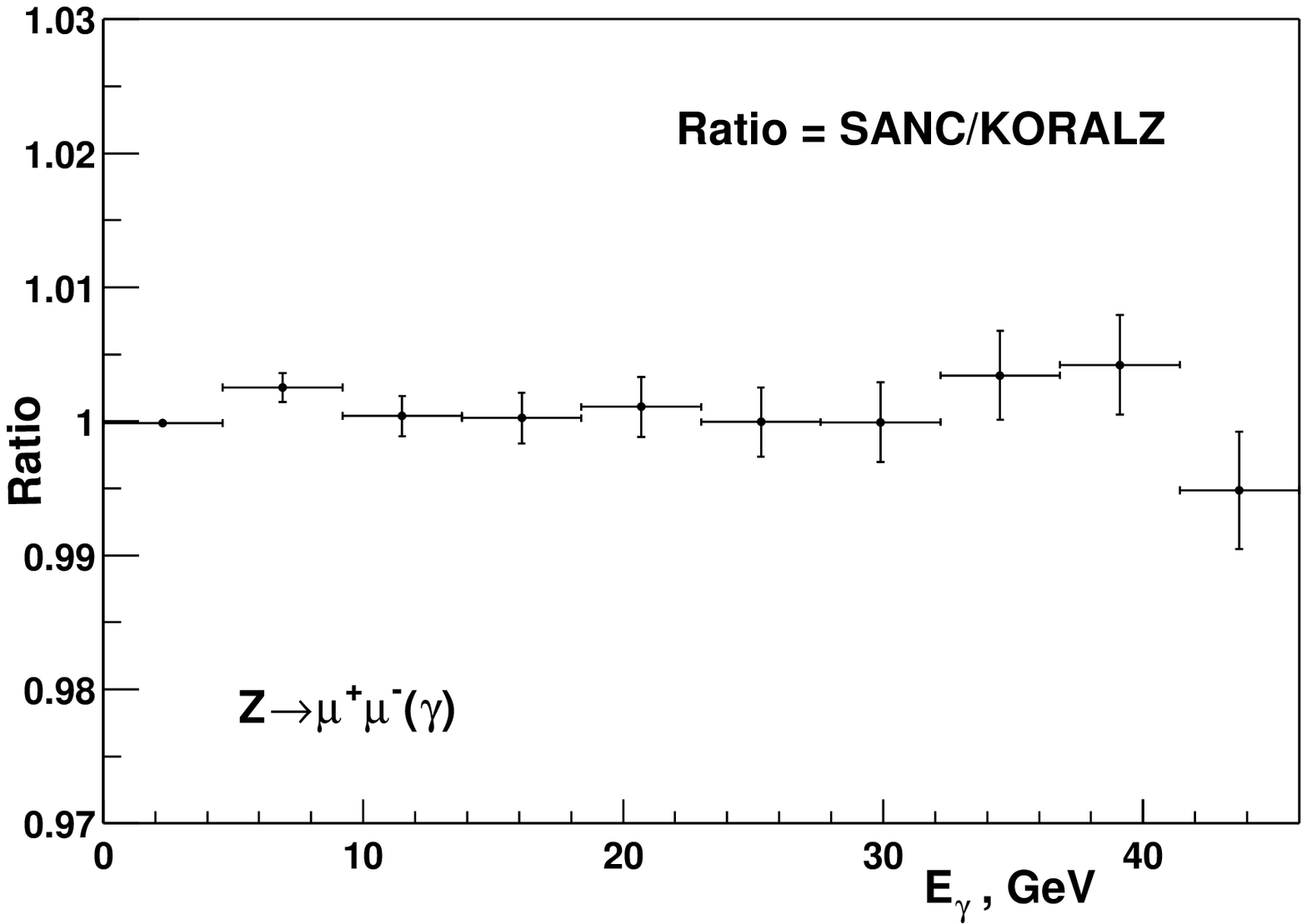,width=80mm,height=65mm}}}  
\put(860, 00){\makebox(0,0)[lb]{\epsfig{file=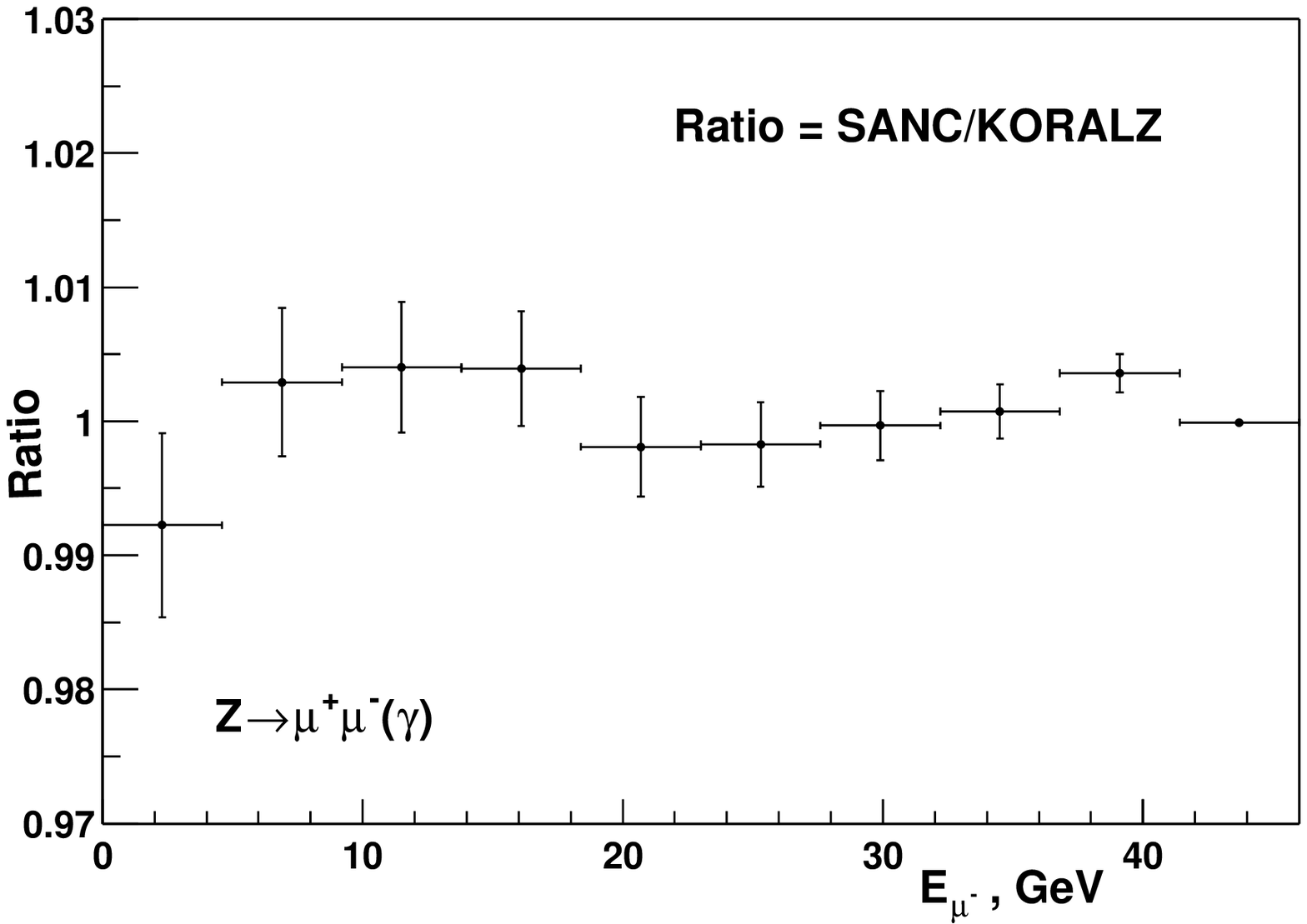,width=80mm,height=65mm}}}  
\end{picture}  
\caption  
{\it Comparisons (ratios) of the \CCol \, and {\tt KORALZ} predictions 
for the $Z$ decay. Observables {\bf A} and {\bf B}: ratios of the photon energy (left-hand side) 
and muon energy (right-hand side)  distributions from the two programs.
The dominant contribution is of leading-log (collinear) nature.}  
\label{figAB-Z}  
\end{figure}  

\begin{figure}[!ht]  
\setlength{\unitlength}{0.1mm}  
\begin{picture}(1600,800)  
\put( 375,750){\makebox(0,0)[b]{\large }}  
\put(1225,750){\makebox(0,0)[b]{\large }}  
\put( -60, 00){\makebox(0,0)[lb]{\epsfig{file=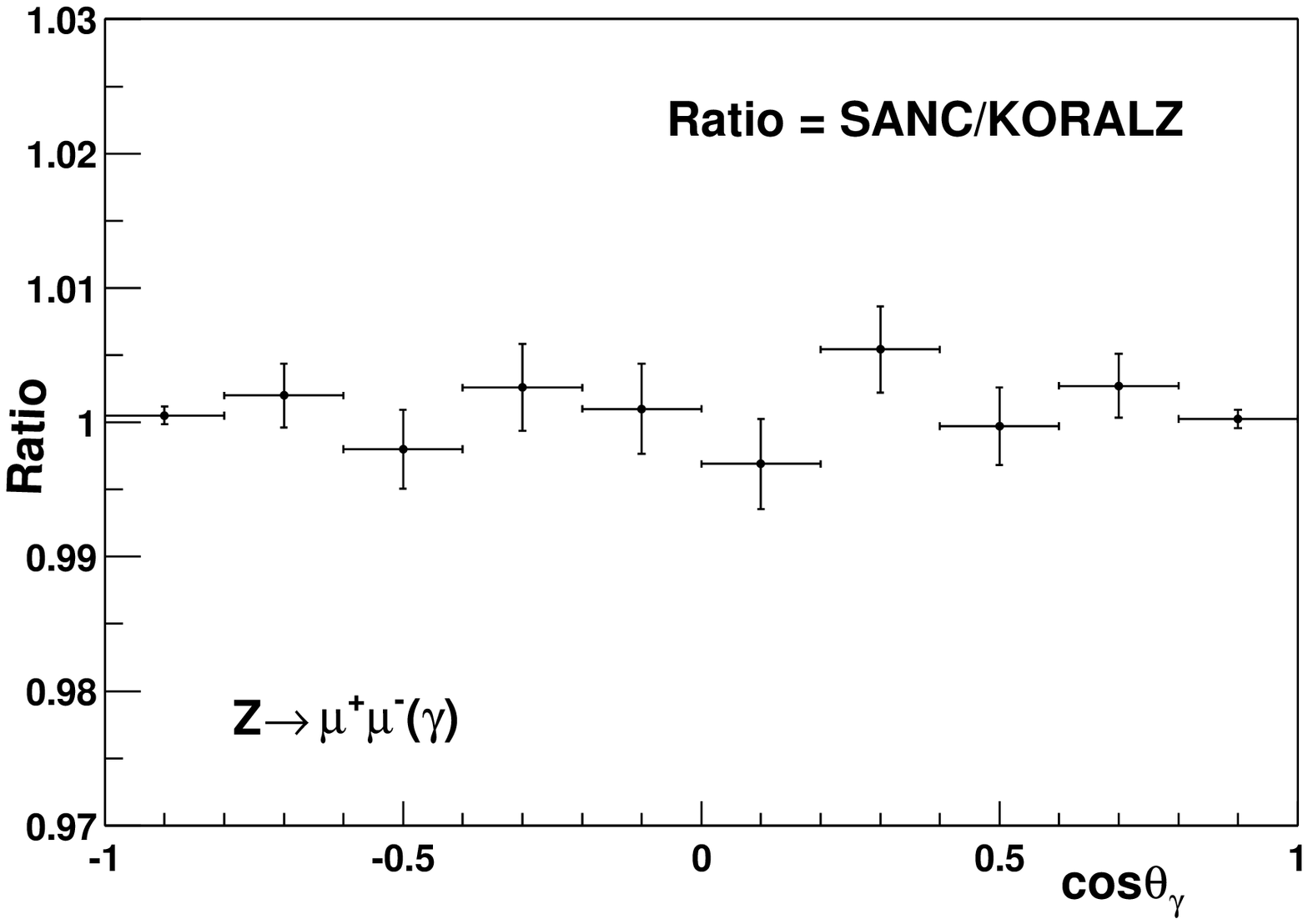,width=80mm,height=65mm}}}  
\put(860, 00){\makebox(0,0)[lb]{\epsfig{file=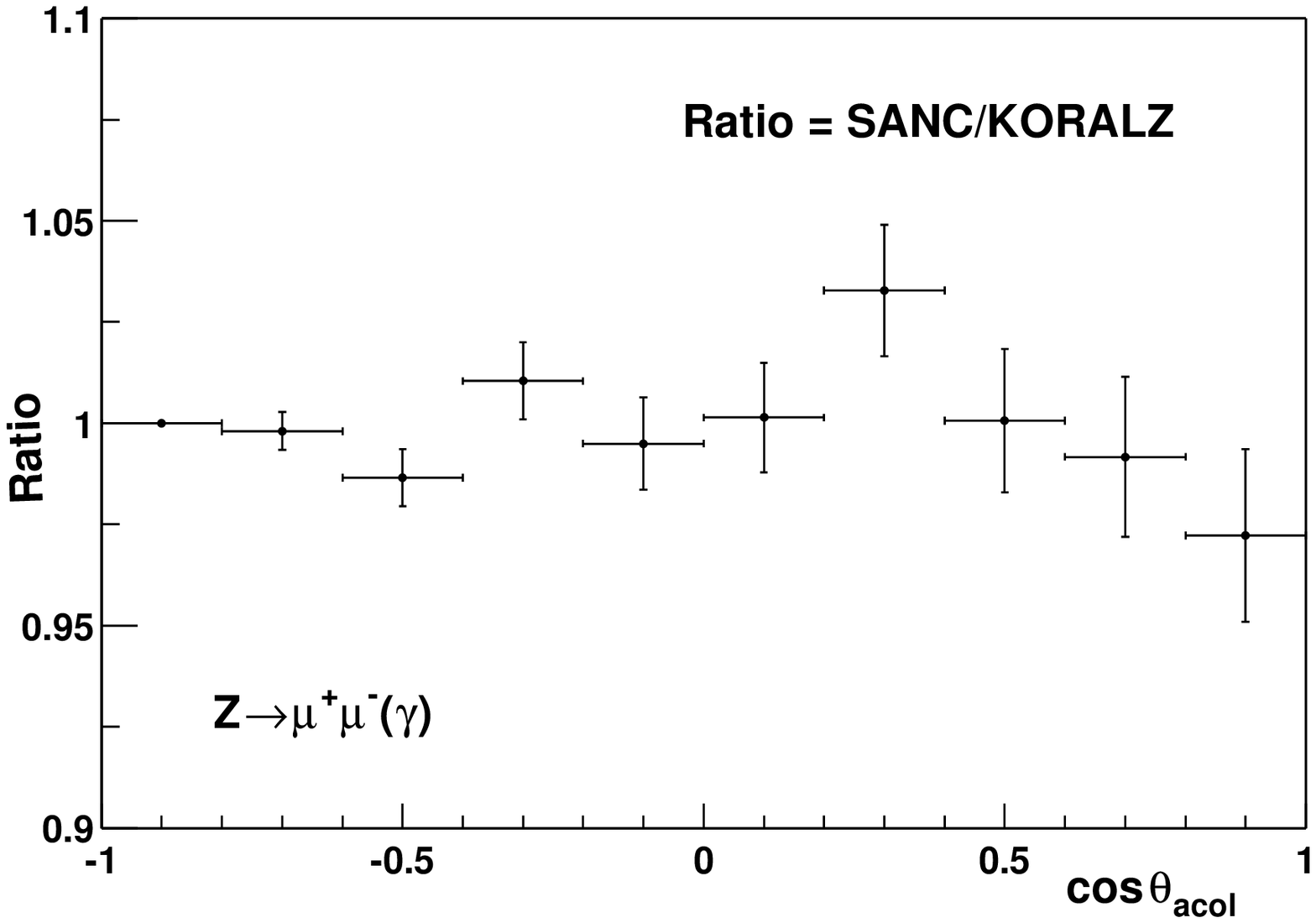,width=80mm,height=65mm}}}  
\end{picture}  
\caption  
{\it Comparisons (ratios) of the \CCol\, and {\tt KORALZ} predictions 
for the $Z$ decay. Observables {\bf C} and {\bf D}: ratios of the photon angle with respect 
to $\mu^-$ (left-hand side) 
and $\mu^-\mu^+$ acollinearity (right-hand side)  distributions from the two programs.
The dominant contribution is of infrared non-leading-log  nature for the left-hand side plot,
and non-infrared non-leading-log nature for the right-hand side one.} 

\label{figCD-Z}  
\end{figure}  

\begin{figure}[!ht]  
\setlength{\unitlength}{0.1mm}  
\begin{picture}(1600,800)  
\put( 375,750){\makebox(0,0)[b]{\large }}  
\put(1225,750){\makebox(0,0)[b]{\large }}  
\put( -60, 00){\makebox(0,0)[lb]{\epsfig{file=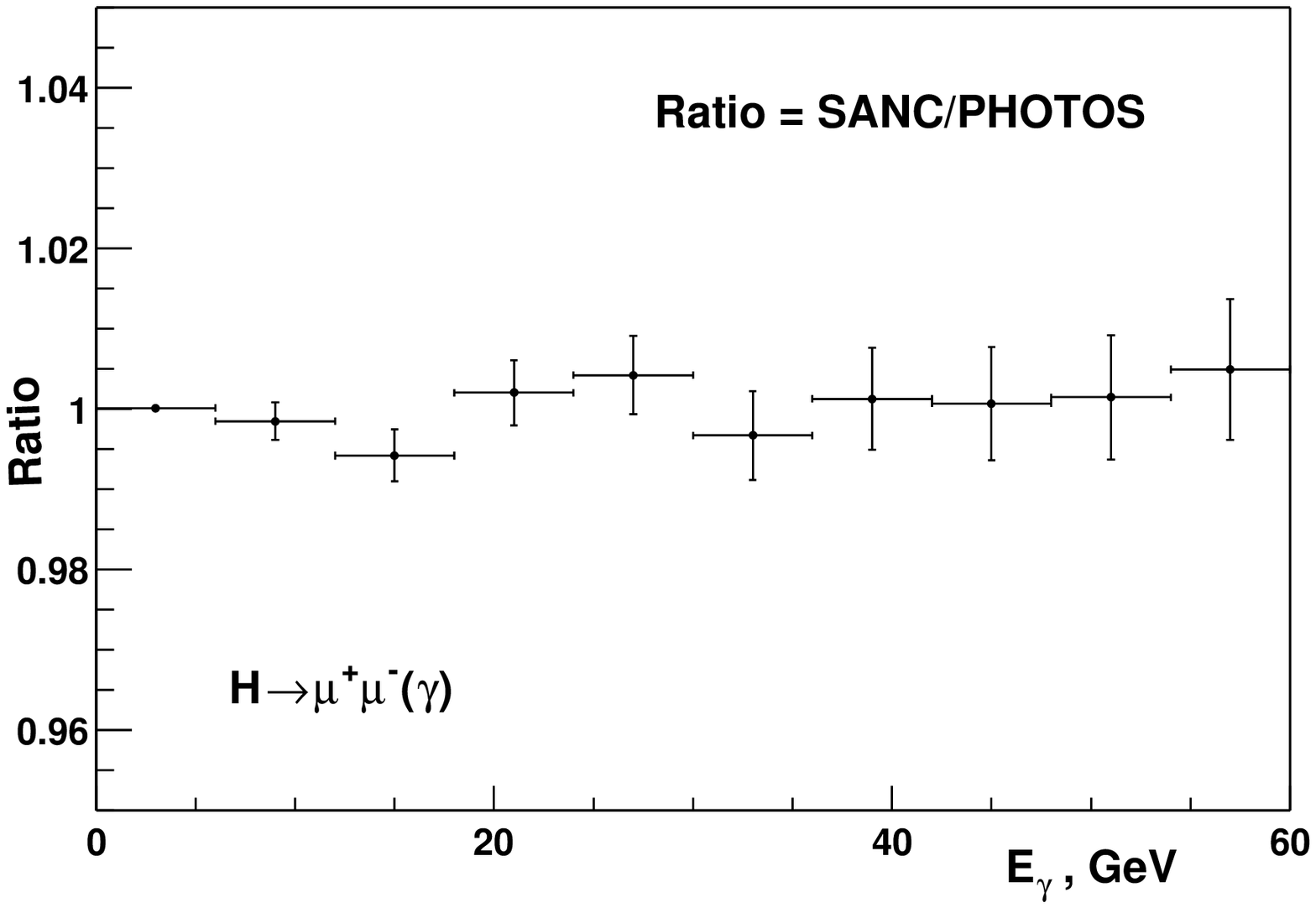,width=80mm,height=65mm}}}  
\put(860, 00){\makebox(0,0)[lb]{\epsfig{file=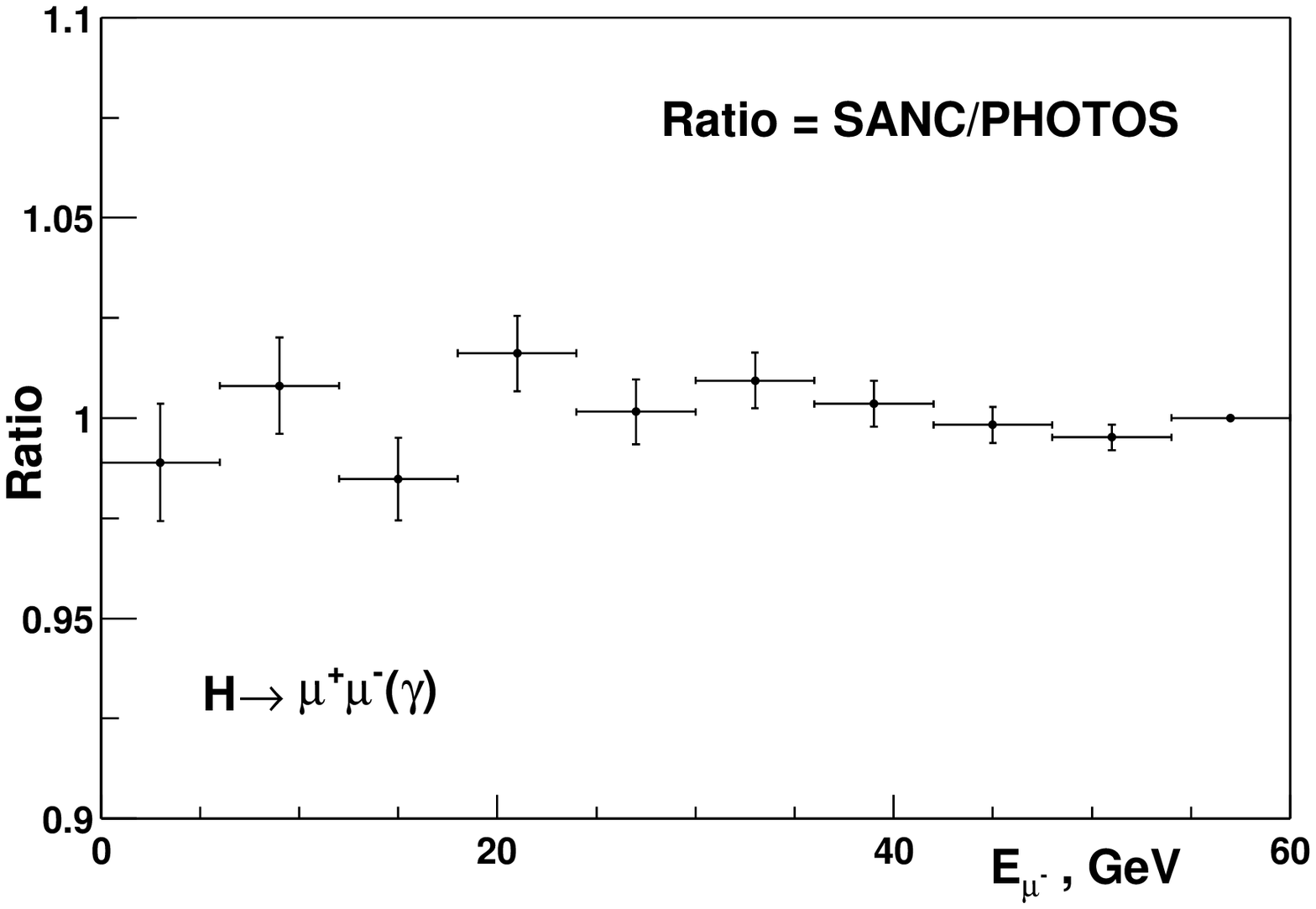,width=80mm,height=65mm}}}  
\end{picture}  
\caption  
{\it Comparisons (ratios) of the \CCol\, and {\tt PHOTOS} predictions 
for the $H$ decay. Observables {\bf A} and {\bf B}: ratios of the photon energy (left-hand side) 
and muon energy (right-hand side)  distributions from the two programs.
The dominant contribution is of  leading-log (collinear) nature.}  
\label{figAB-H}  
\end{figure}  

\begin{figure}[!ht]  
\setlength{\unitlength}{0.1mm}  
\begin{picture}(1600,800)  
\put( 375,750){\makebox(0,0)[b]{\large }}  
\put(1225,750){\makebox(0,0)[b]{\large }}  
\put( -60, 00){\makebox(0,0)[lb]{\epsfig{file=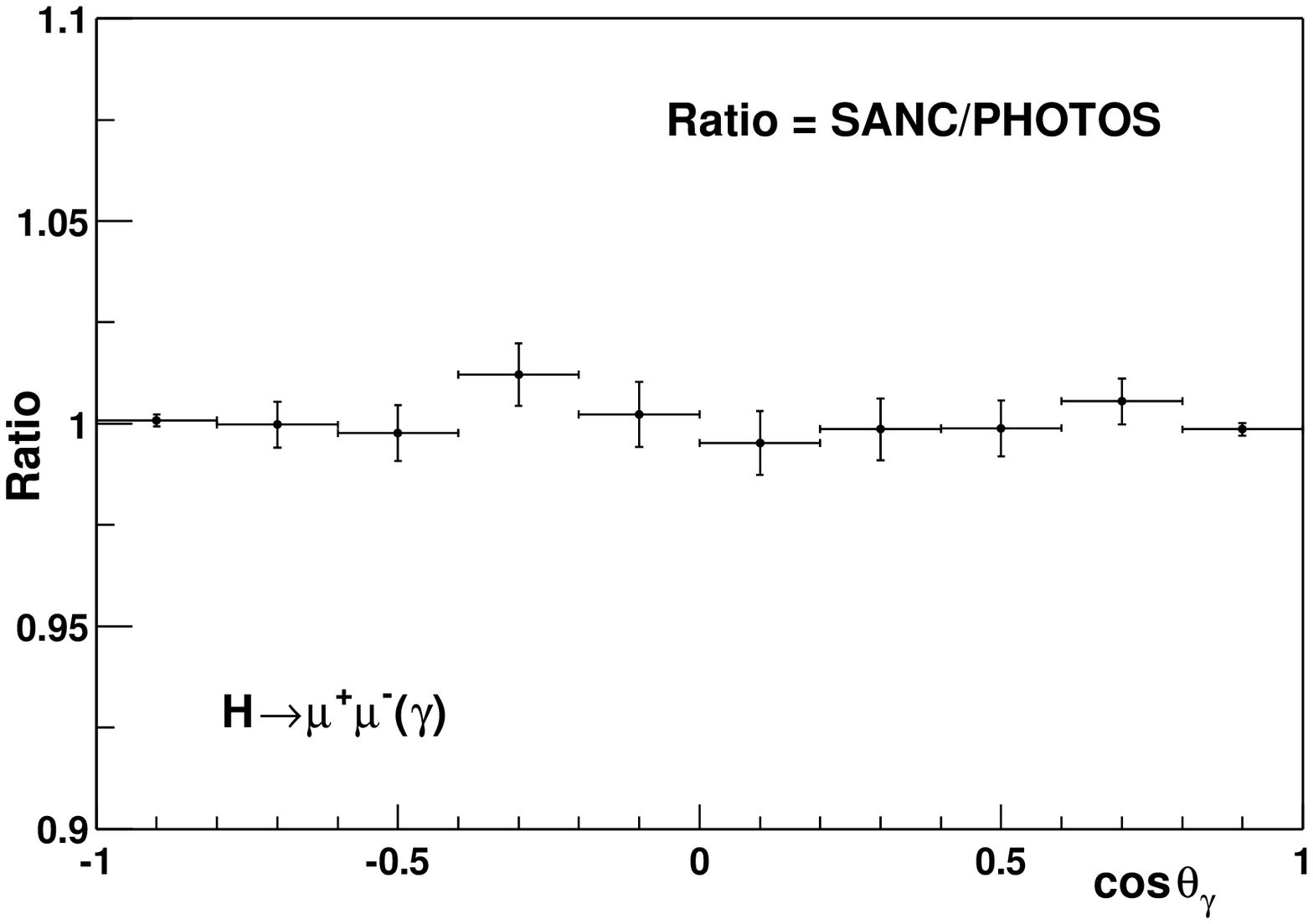,width=80mm,height=65mm}}}  
\put(860, 00){\makebox(0,0)[lb]{\epsfig{file=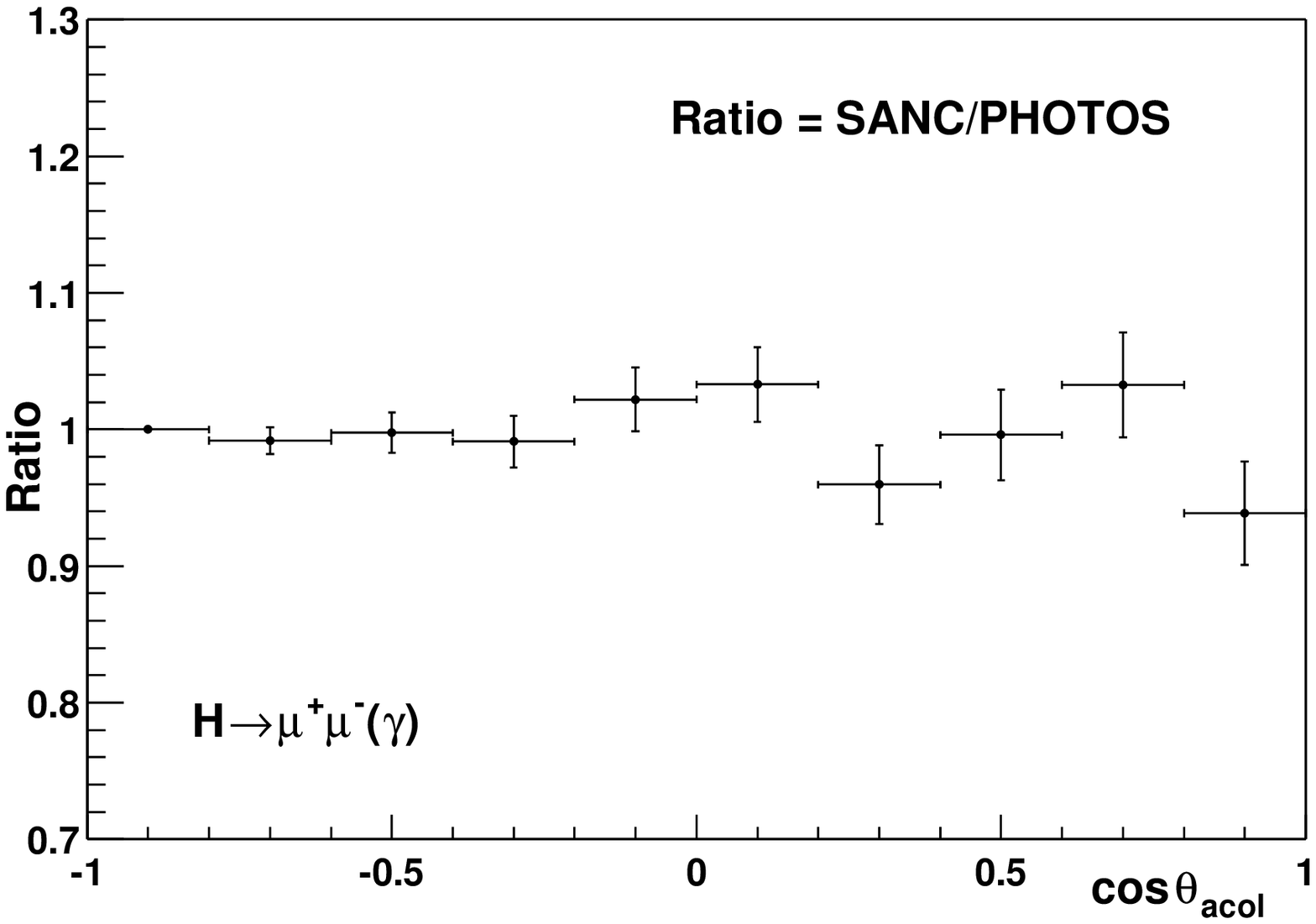,width=80mm,height=65mm}}}  
\end{picture}  
\caption  
{\it Comparisons (ratios) of the \CCol\, and {\tt PHOTOS} predictions 
for the $H$ decay. Observables {\bf C} and {\bf D}: ratios of the photon angle with respect 
to $\mu^-$ (left-hand side) 
and $\mu^-\mu^+$ acollinearity (right-hand side)  distributions from the two programs. The
dominant contribution is of infrared non-leading-log  nature for the left-hand side plot,
and non-infrared non-leading-log nature for the right-hand side one.} 
\label{figCD-H}  
\end{figure}  

\begin{figure}[!ht]  
\setlength{\unitlength}{0.1mm}  
\begin{picture}(1600,800)  
\put( 375,750){\makebox(0,0)[b]{\large }}  
\put(1225,750){\makebox(0,0)[b]{\large }}  
\put( -60, 00){\makebox(0,0)[lb]{\epsfig{file=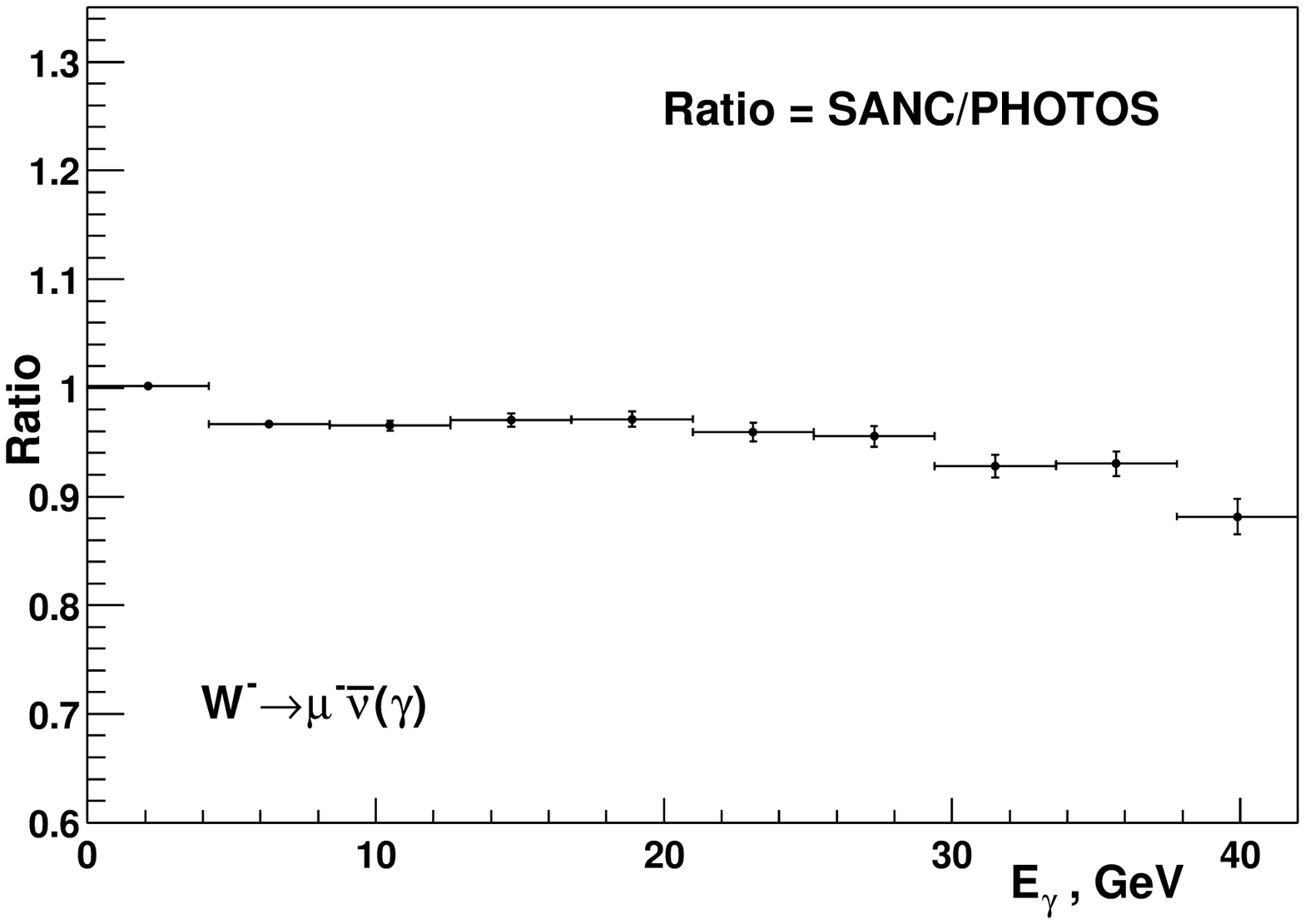,width=80mm,height=65mm}}}  
\put(860, 00){\makebox(0,0)[lb]{\epsfig{file=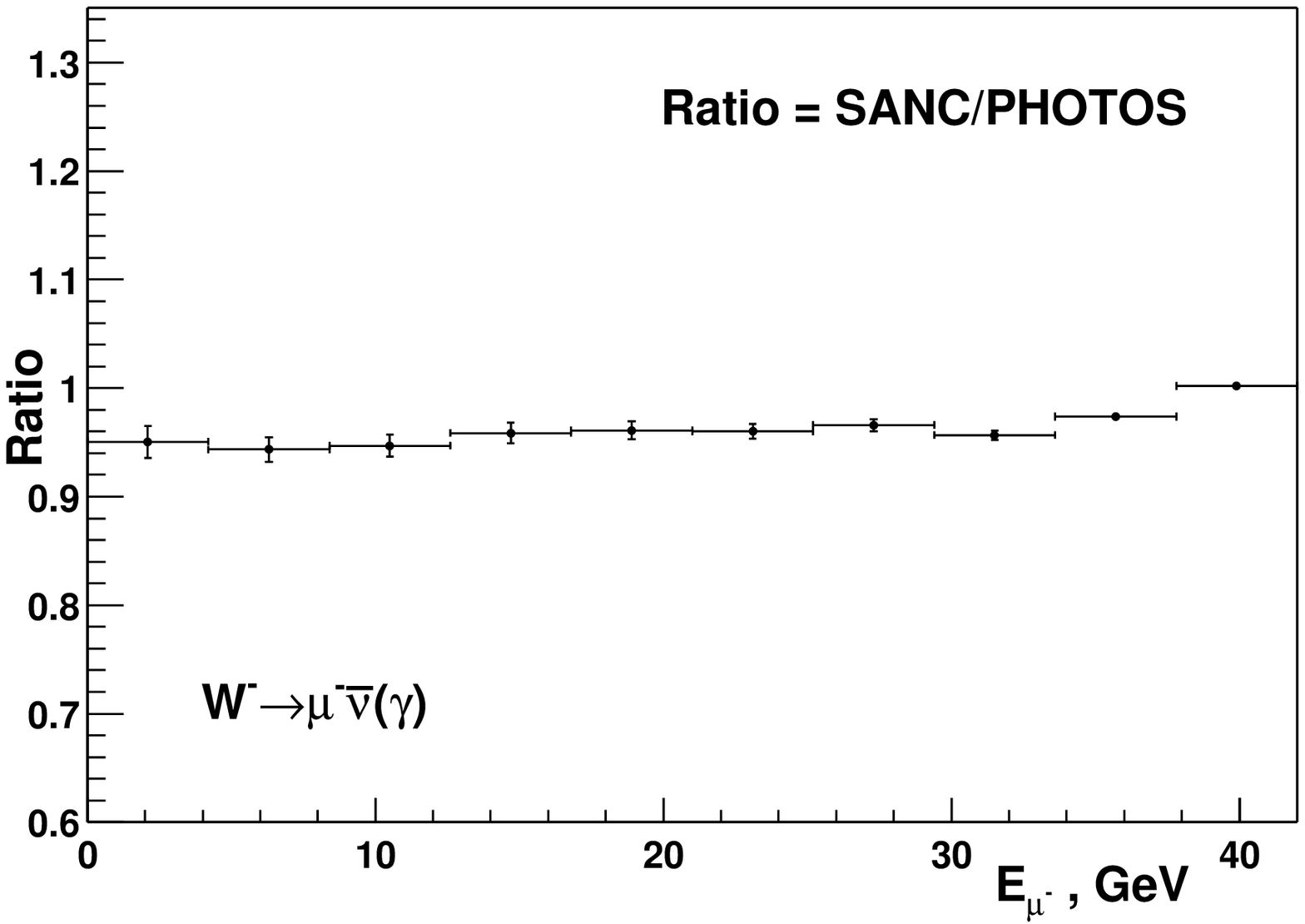,width=80mm,height=65mm}}}  
\end{picture}  
\caption  
{\it Comparisons (ratios) of the \CCol\,and {\tt PHOTOS} predictions 
for the $W$ decay. Observables {\bf A} and {\bf B}: ratios of the photon energy (left-hand side) 
and muon energy (right-hand side)  distributions from the two programs. The
dominant contribution is of leading-log (collinear) nature.}  
\label{figAB-W}  
\end{figure}  

\begin{figure}[!ht]  
\setlength{\unitlength}{0.1mm}  
\begin{picture}(1600,800)  
\put( 375,750){\makebox(0,0)[b]{\large }}  
\put(1225,750){\makebox(0,0)[b]{\large }}  
\put( -60, 00){\makebox(0,0)[lb]{\epsfig{file=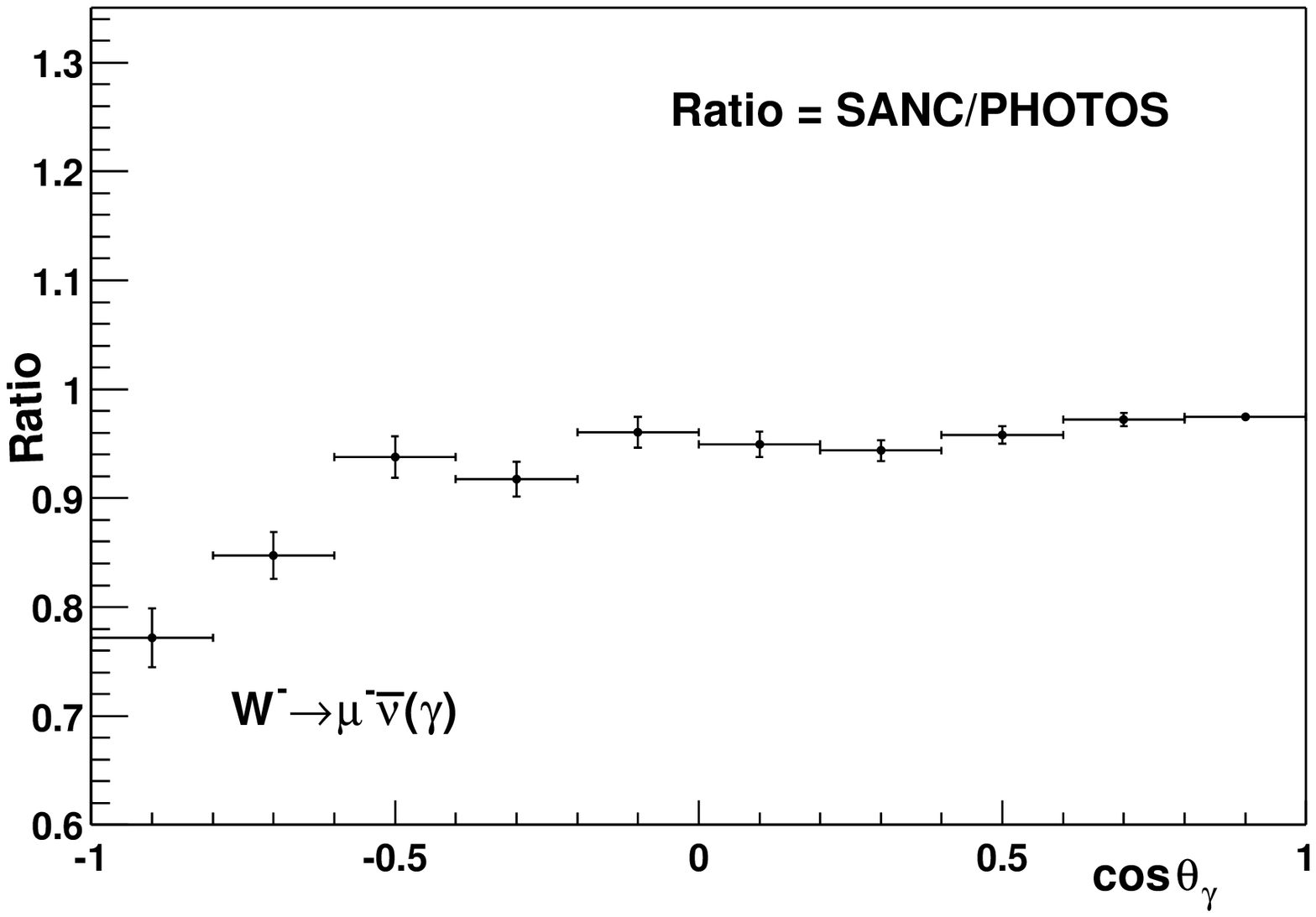,width=80mm,height=65mm}}}  
\put(860, 00){\makebox(0,0)[lb]{\epsfig{file=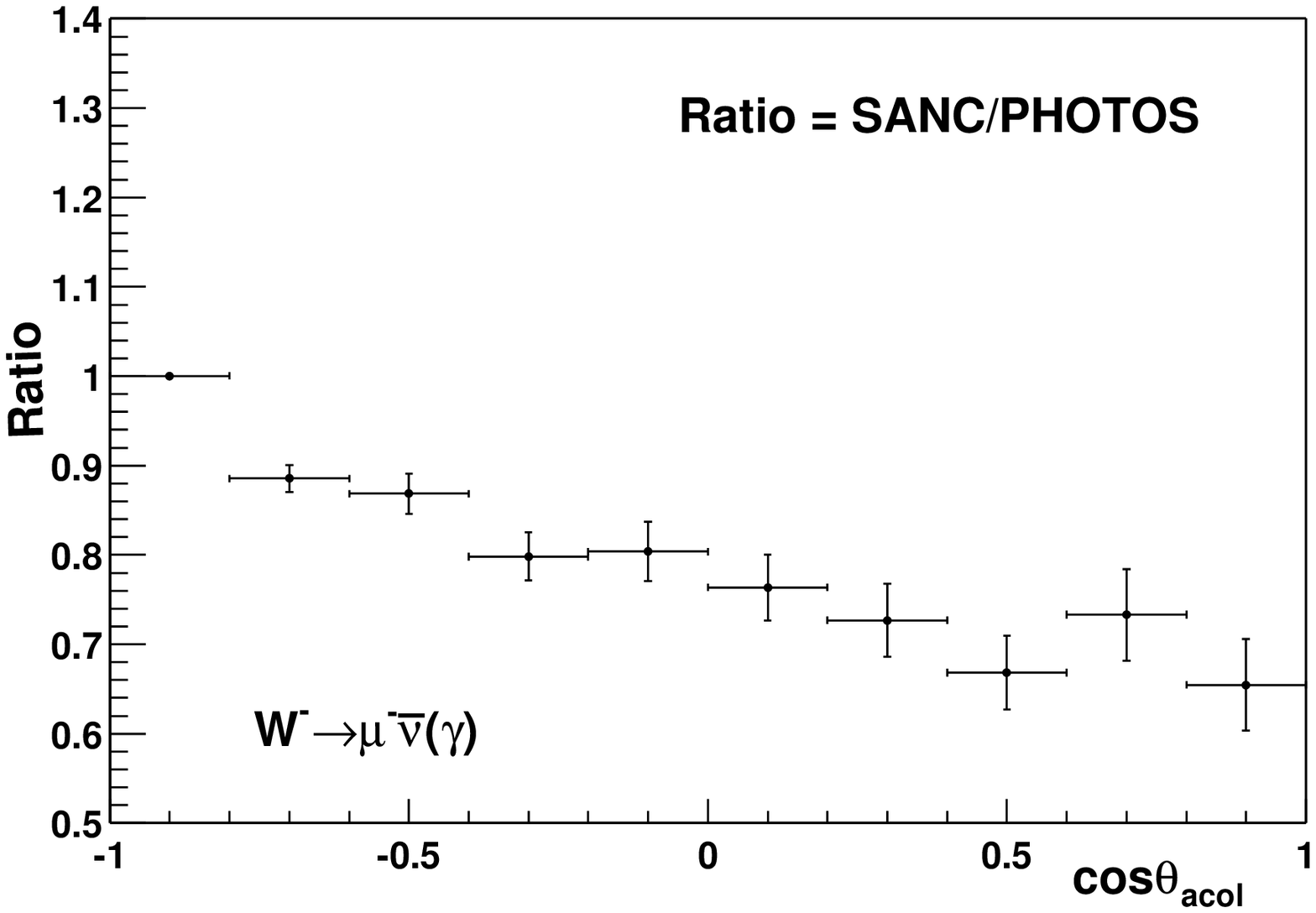,width=80mm,height=65mm}}}  
\end{picture}  
\caption  
{\it Comparisons (ratios) of the \CCol\, and {\tt PHOTOS} predictions 
for the $W$ decay. Observables {\bf C} and {\bf D}: ratios of the photon angle with respect 
to $\mu^-$ (left-hand side) 
and $\mu^-\mu^+$ acollinearity (right-hand side)  distributions from the two programs. The
dominant contribution is of  infrared non-leading-log  nature for the left-hand side plot,
and non-infrared non-leading-log nature for the right-hand side one.} 
\label{figCD-W}  
\end{figure}  

In the case of the $W$ and Higgs decay we rely on the comparison
with {\tt PHOTOS}. Due to the incomplete one-loop QED
in {\tt PHOTOS} we cannot, in priciple,
expect the agreement for the decay distributions {\bf A, B} to be better than 
the order of ${1 \over \ln{m_B^2/m_\mu^2 }} \simeq $  7\%.
We see in fig. 3 for Higgs boson decay and in fig. 5  
for $W$ decay, that this  is indeed the case.
In fact, the agreement is much better in the case of the Higgs boson decay.
In figs. 4 and 6 we see the angular distributions of the photon with respect 
to the charged fermion (observable {\bf C}).
Here, in principle, the agreement should
not be worse than ${1 \over \ln{m_B/E_\gamma^{min} }} \simeq $  20\%
for the photon emitted in directions far from the charged particle
and  ${1 \over {\ln{m_B^2/m_\mu^2 } \ln{m_B/E_\gamma^{min} }}} \simeq $  1.4\%
for directions close to the charged particle.
This is indeed the case.

\section{Physical uncertainties of {\tt PHOTOS} in $W$ and $H$ decay }

In the case of testing $H$ and $W$ decays, the comparisons of {\tt PHOTOS}
with the ``matrix-element'' type calculations
were never accounted for in a well documented way.
The following discussion will give us a chance systematically  evaluate
uncertainties of {\tt PHOTOS} for these two decays.

In the case of the Linear Collider studies and  many scenarios for new physics,
the cross section for the production of the Higgs boson can be quite large. 
In the case of sophisticated observables, the  reconstruction 
of the reference frames of the intermediate states requires good control
of the bremsstrahlung corrections.
Our comparison shows that, in the case of the Higgs boson decay,
the approximation used in {\tt PHOTOS}  works much better than could be expected
from its design principles.
In fact it provides results difficult to distinguish from the 
``matrix-element'' ones in the case of all observables {\bf A} to {\bf D}
(figs. 3 and 4).

At LEP2, the production and decay of $W$  pairs   is now being combined
for all four LEP2 experiments, and
uncertainties due to bremsstrahlung in $W$ decay are important.
{\tt PHOTOS} is part of one of the main programs (YFSWW3)
used in the LEP2 analysis of the  $W$-pair data.
Our tests will provide an estimate of the size of the uncertainties, due to use of
{\tt PHOTOS}.
If they turn out to be sizeable, the improvements
will need to be implemented either into the  {\tt PHOTOS} algorithm or by other means.

In figs. 5 and 6, we see that up to  leading order,
the distributions agree with the ``matrix element'' results provided by  \CCol.
There is, however, no exceptionally 
good agreement observed in the previous case of $H$ decay.

This point  requires further study,
before planning improvements of  {\tt PHOTOS}.
In particular, a comparison of {\tt PHOTOS} with another
available matrix-element calculation of $\tau \to e \nu \bar \nu (\gamma)$,
e.g. from {\tt TAUOLA } \cite{tauola:1992}, should also be repeated to check 
if a pattern of discrepancy  similar to that for $W$ decay is observed.
The $W$ channel  is of importance for some studies 
\cite{Richter-Was:1993ta} of LHC Higgs discovery potential as well, 
see also \cite{ATLASTDR}.

\section{Summary}

We have successfully tested \CCol\ versus {\tt KORALZ}
in case of the $Z$ decay.
In this way we have verified the technical correctness of {\tt SANC},
e.g. its phase-space generation and QED correction amplitudes.

In case of $W$ and $H$ decays,
we have checked that \CCol\ agrees with the leading order QED calculation
provided by the {\tt PHOTOS} Monte Carlo. 
These comparisons allow us to evaluate the size
of missing non-leading terms in {\tt PHOTOS}.
In the case of $H$ decays, we have found that {\tt PHOTOS}
results are exceptionally  good -- differences with \CCol\ are indistinguishable 
from zero and below 1\% everywhere.
In the case of $W$ decay, {\tt PHOTOS} predictions are within 7\% for the 
end parts of the spectra affected by the leading-log corrections and
within 20\% for the angular part of the distributions, where the 
infrared-induced logarithm dominates over non-infrared non-leading terms only.
The differences are up to 40\% in the phase-space  regions 
where only non-leading corrections contribute to the matrix element. 
We can conclude that results from all three programs for
all four observables agree within the
expected precision ranges.

\section*{Acknowledgements}
The authors would like to thank D. Bardin and B.F.L. Ward
for useful discussions.

\newpage


\end{document}